\documentclass[sigconf,screen]{acmart}

\AtBeginDocument{%
  }

\usepackage{graphicx}
\usepackage{subcaption}
\usepackage{adjustbox}
\usepackage{xcolor}
\usepackage{xspace}

\newcommand{\todo}[1]{{\color{red}{\textbf{#1}}}}

\newcommand{\authortodo}[2]{\todo{#1: #2}}

\newcommand{\pat}[1]{\authortodo{Pat}{#1}}
\newcommand{\pmr}[1]{\pat{#1}}

\newcommand{\mt}[1]{\authortodo{Michele}{#1}}
\newcommand{\samcheng}[1]{\authortodo{Sam}{#1}\xspace}

\newcommand{\brt}{BRT\xspace}
\newcommand{\brtfull}{Bug Reproduction Test\xspace}
\newcommand{\libro}{LIBRO\xspace}
\newcommand{\llm}{LLM\xspace}
\newcommand{\tool}{BRT Agent\xspace}
\newcommand{\react}{ReAct\xspace}
\newcommand{\passerine}{Passerine\xspace}
\newcommand{\failtoany}{$F \rightarrow X$\xspace}
\newcommand{\failtopass}{$F \rightarrow P$\xspace}
\newcommand{\autoprfull}{Automated Program Repair\xspace}
\newcommand{\autopr}{APR\xspace}
\newcommand{\enpassratefull}{Ensemble Pass Rate\xspace}
\newcommand{\enpassrate}{EPR\xspace}
\newcommand{\irfull}{Information Retrieval\xspace}

\newcommand{\defectsfj}{Defects4J\xspace}
\newcommand{\github}{GitHub\xspace}
\newcommand{\swebench}{SWE-Bench\xspace}
\newcommand{\swtbench}{SWT-Bench\xspace}
\newcommand{\sweagent}{SWE-Agent\xspace}
\newcommand{\sweagentplus}{\sweagent{}+\xspace}
\newcommand{\chatgpt}{ChatGPT\xspace}
\newcommand{\codex}{Codex\space}
\newcommand{\gptthreepfive}{GPT-3.5\xspace}
\newcommand{\aider}{AIDER\xspace}
\newcommand{\autocoderover}{AutoCodeRover\xspace}
\newcommand{\fixpatch}{fix\xspace}

\newcommand{\google}{Google\xspace}
\newcommand{\gemini}{Gemini\xspace}
\newcommand{\gits}{GITS\xspace}

\newcommand{\actioncat}{\CodeIn{cat}\xspace}
\newcommand{\actioncodesearch}{\CodeIn{code\_search}\xspace}
\newcommand{\actiontest}{\CodeIn{bazel test}\xspace}
\newcommand{\actionfinish}{\CodeIn{finish}\xspace}
\newcommand{\actionedit}{\CodeIn{edit}\xspace}
\newcommand{\codeeditingllm}{code-editing \llm{}\xspace}

\newcommand{\rqonetitle}{\brt Generation Effectiveness\xspace}
\newcommand{\rqtwotitle}{Impact of \brt on Fix Generation\xspace}
\newcommand{\rqthreetitle}{Impact of \brt on Fix Selection\xspace}

\newcommand{\Space}[1]{} %
\newcommand{\Comment}[1]{}
\newcommand{\CodeIn}[1]{\begin{small}\texttt{#1}\end{small}}
\newcommand{\myparagraph}[1]{\paragraph{#1}}

\acmBooktitle{Proceedings of ACM Conference (Conference’17)}

\begin{document}

\title{Agentic Bug Reproduction for Effective Automated Program Repair at Google}

\author{Runxiang Cheng$^{1*}$, Michele Tufano$^{2}$, Jürgen Cito$^{3*}$, José Cambronero$^{2}$\\ Pat Rondon$^{2}$, Renyao Wei$^{2}$, Aaron Sun$^{2}$, Satish Chandra$^{2}$}
\thanks{*Runxiang Cheng and Jürgen Cito conducted this research at Google.}
\affiliation{%
  \institution{$^{1}$University of Illinois Urbana-Champaign\\ $^{2}$Google\\
  $^{3}$TU Wien, Austria}
  \country{}
}

\renewcommand{\shortauthors}{R. Cheng, M. Tufano, J. Cito, J. Cambronero, P. Rondon, R. Wei, A. Sun, and S. Chandra}

\begin{abstract}
Bug reports often lack sufficient detail for developers to reproduce and fix the underlying defects. 
Bug Reproduction Tests (BRTs), tests that fail when the bug is present and pass when it has been resolved, are crucial for debugging, but they are rarely included in bug reports, both in open-source and in industrial settings. 
Thus, automatically generating BRTs from bug reports has the potential to accelerate the debugging process and lower time to repair.
This paper investigates automated BRT generation within an industry setting, specifically at Google, focusing on the challenges of a large-scale, proprietary codebase and considering real-world industry bugs extracted from Google's internal issue tracker. 
We adapt and evaluate a state-of-the-art BRT generation technique, LIBRO, and present our agent-based approach, BRT Agent, which makes use of a fine-tuned Large Language Model (LLM) for code editing. 
Our BRT Agent significantly outperforms LIBRO, achieving a 28\% plausible BRT generation rate, compared to 10\% by LIBRO, on 80 human-reported bugs from Google's internal issue tracker. 
We further investigate the practical value of generated BRTs by integrating them with an Automated Program Repair (APR) system at Google. 
Our results show that providing BRTs to the APR system results in 30\% more bugs with plausible fixes.  
Additionally, we introduce Ensemble Pass Rate (EPR), a metric which leverages the generated BRTs to select the most promising fixes from all fixes generated by APR system.
Our evaluation on EPR for Top-K and threshold-based fix selections demonstrates promising results and trade-offs. 
For example, EPR correctly selects a plausible fix from a pool of 20 candidates in 70\% of cases, based on its top-1 ranking.
\end{abstract}



\maketitle

\section{Introduction}
\label{sec:intro}
Bug reproduction is a critical part of the software debugging process~\cite{chaparro2019assessing,kang2023large,koyuncu2019ifixr,Ernst2014Defects4J,le2012systematic,xiong2017precise,li2019deepfl,yang2024swe}.
\Space{To enable effective and efficient debugging and repairing, b}The bug reproduction steps of a reported bug can often be encapsulated into test(s), which automatically reproduce the bug by demonstrating test failure(s).
Such a test is referred to as a \brtfull (\brt).
More formally, a \brt\Space{ is a test case\samcheng{do we use both 'test case' and 'test' in the paper? if not, can just call test.}} should (1) fail in the presence of the bug, demonstrating the faulty behavior; and (2) pass once the bug has been resolved, confirming the effectiveness of the applied fix patch.
For both human users and \autoprfull (\autopr) systems, \brt{}s can offer insights on the root cause of the bug~\cite{ko2008debugging,kang2023large,nayrolles2015jcharming,just2018comparing}, \Space{plays a crucial role in guiding}guide the development of a fix patch~\cite{rondon2025passerine,yang2024swe,mundler2024swt}, and is essential for validating the effectiveness of the proposed fix patch~\cite{ko2008debugging,Ernst2014Defects4J,saha2017elixir,le2012systematic}.

However, despite their importance, \brt{}s are often absent\Space{ from bug reports} when the debugging or repairing process starts.
Bug reports from human users/developers, written in natural language, are often vague, ambiguous, and lacking crucial details for reproducing and subsequently addressing the reported software defects~\cite{bettenburg2008duplicate,zimmermann2009cross,anvik2006should}.
Studies show that bug reports in open-source projects often do not have \brt{}s, e.g., \citet{koyuncu2019ifixr} found that only 4\% of the bug reports in \defectsfj~\cite{Ernst2014Defects4J} include a \brt; \citet{kang2023large} found that 28\% of the tests among the 300 most-starred \github projects are added due to bug reports;
more recently, \citet{mundler2024swt} found that there are no corresponding \brt{}s for bug-reporting pull requests prior to the creation of those pull requests in \swebench projects.

In industrial setting, similarly, \brt{}s are commonly expected to be developed \textit{alongside} the fix patch, after the bug is reported.
Through our exploration of \brt creation and \brt availability in \google's internal issue tracking system\Space{ (\gits)}, we observed that \brt development can be deferred to fix development stage because developers often do not prioritize the task of, or possess enough information of, developing a \brt at the time of bug reporting (\S\ref{sec:motivation}).
For example, developers may not attend to \brt development when actively dealing with issues reported by production monitoring.
Moreover, developing a \brt requires knowledge of the bug's root cause, and effort to craft a precise, bug-reproducing test failure~\cite{grottke2007fighting,just2018comparing,beller2018dichotomy}, which may not be available\Space{ immediately} as soon as the bug is reported. 

In this paper, we tackle the challenge of automated \brt generation in an industrial\Space{ production} environment, specifically at \google.
Automated \brt generation facilitates \brt development at scale, which could provide crucial benefits to industrial settings where a large number of bug reports are handled per day.

To develop automated \brt generation for \google's\Space{ internal} development environment, we first make nontrivial effort to adapt a state-of-the-art \llm-based approach, \libro~\cite{kang2023large} (\S\ref{sec:approach:libro}).
After finding limitations from the adapted \libro, we developed our \llm-agent-based approach for \brt generation, \tool (\S\ref{sec:approach:agent})\Space{, independently and concurrently to a recently proposed \llm-based \brt generation agent~\cite{mundler2024swt} (\S\ref{sec:relatedwork:compare})}.
Our approach, \tool, goes beyond being a straightforward adaptation of the existing\Space{ open-source} \llm-agent-based \autopr systems into \brt generation task via \llm prompt engineering (\S\ref{sec:relatedwork}).
Specifically, we introduce and focus on agent components that overcome challenges unique to our industrial context. 
For example, to address the challenge of generating proprietary code with idiosyncratic APIs and coding practices, \tool incorporates a specialized \codeeditingllm fine-tuned on \google's codebase, to generate high-quality test code (\S\ref{sec:approach:agent:action}).
We implemented both \libro and our \tool to work with \google's development infrastructure, and evaluate them on a set of 80 production bugs reported and fixed by human developers from a wide array of \google projects~\cite{rondon2025passerine} (\S\ref{sec:empiricalstudy:datasetandconfig}).

Moreover, while prior work focuses on evaluating the performance of \brt generation techniques in isolation (\S\ref{sec:relatedwork}), we further investigate the practical impact of the generated \brt{}s for\Space{ the end-to-end performance of automated} bug-fixing.
Given \brt{}s can provide additional context for understanding the bug, we study how generated \brt{}s can improve the performance of \passerine~\cite{rondon2025passerine}, an industrial-scale \autopr system at \google, to generate plausible fixes for more bugs more efficiently. 
Given\Space{ recent} \autopr systems can generate many fixes per bug and \brt{}s can help validate a fix through Fail-to-Pass behavior (\S\ref{sec:motivation}), we explore whether generated \brt{}s can be used to effectively indicate the plausible fixes from all fixes generated by \passerine\Space{, for bugs where no existing tests had failed}. 
To this aim, we propose a novel metric, \enpassratefull (\enpassrate), which evaluates each generated fix by calculating its pass rate over a suite of generated \brt{}s, to select promising fixes from \autopr systems. 

This paper makes the following key contributions:
\begin{itemize}
    \item We adapt state-of-the-art approach, \libro, and develop our \llm-agent-based approach for \brt generation, \tool, in \google's environment (\S\ref{sec:approach}).
    \tool substantially outperforms \libro by generating plausible \brt{}s on 28\% of the evaluated \google production bugs across six programming languages, compared to that of 10\% by \libro.
    Our manual inspection find that 67\% of the plausible \brt{}s generated by \tool are semantically equivalent or identical to the oracle \brt in the ground truth fixes; we also provide insights to the agent behaviors during \brt generation (\S\ref{sec:results:brt}).
    \item We assess the practical impact of generated plausible \brt{}s on improving \passerine's fix generation performance.
    When generated \brt{}s are provided as part of the bug report, \passerine fixes 30\% more bugs, and fewer agent steps were taken to generate a fix (\S\ref{sec:results:generation}).
    \item We propose \enpassratefull (\enpassrate), a metric using generated \brt{}s to help select the more promising fixes from all \autopr-generated fixes for bugs where no existing test had failed. Our evaluation of \enpassrate on Top-K fix selection and threshold-based fix selection shows promising results and reveals trade-offs (\S\ref{sec:results:selection}). 
\end{itemize}

\section{Background and Motivation}
\label{sec:motivation}
 
In this section, we describe the definition of \brt (\S\ref{sec:movtivation:def}), the availability of \brt in the industrial setting (\S\ref{sec:movtivation:available}), and the need for automated \brt generation (\S\ref{sec:movtivation:need}).

\subsection{\brt Definition}
\label{sec:movtivation:def}

A \brt is a test that exhibits a Fail-to-Pass (\failtopass) behavior: the test fails when executed against a buggy codebase, but passes when executed against the fixed codebase~\cite{mundler2024swt}. 
The implementation of the \brt is useful for understanding and localizing the reproduced bug, aiding the development of an effective fix patch to the bug, while the \failtopass behavior of \brt provides necessary evidence that the proposed fix patch to the reproduced bug is correct~\cite{ko2008debugging,kang2023large,mundler2024swt,yang2024swe,nayrolles2015jcharming,rondon2025passerine,Ernst2014Defects4J,saha2017elixir,koyuncu2019ifixr,le2012systematic}. 
Program repair literature often denotes ``fix patch'' as ``patch''; however, to avoid ambiguity with the test patch in \brt development, we denote \textit{fix patch} for a bug as \textit{fix}.

\subsection{\brt Availability}
\label{sec:movtivation:available}

Prior studies have shown that \brt{}s are scarce at time of bug reporting in open-source projects~\cite{koyuncu2019ifixr, mundler2024swt} (\S\ref{sec:intro}).
In the industrial setting, the common expectation is that \brt{}(s) are developed alongside the fix\Space{, after the bug report is filed}\Space{, transitioning from a failing to a passing state (\failtopass)}\footnote{A \brt could be available as evidence in the bug report from certain \textit{machine-reported} bug categories, e.g., bugs reported by dynamic analyses like AddressSanitizer~\cite{addresssanitizer}.}.
The practice of deferring \brt development to the fix development stage stems from the fact that bug reports originate from sources that often do not readily prioritize the task of, or possess the knowledge of, developing \brt{}s at the time of bug reporting. 

Specifically, typical bug report sources are team-internal developers, team-external developers, and end users. 
Team-internal developers understand the codebase but may not prioritize writing a \brt immediately when encountering an issue, particularly during active development or when dealing with issues surfaced by production monitoring systems. 
Team-external developers understand the system's expected functionality but may not have the expertise to write a bug-targeted test.
End users have limited knowledge of the underlying code, and focus primarily on describing the observed issue\Space{ instead of writing tests}. 
It is also important to note that issues reported by end users are typically not classified as internal bugs in our context.\Comment{\samcheng{why is this note relevant?} \mt{the origin of this note was to highlight the fact that in our study we do not consider user-submitted bug reports, but only internal bugs, i.e., Googlers submitted bugs.}}

Regardless of the bug report source's exposure to the codebase, writing an effective \brt is  challenging and time-consuming.
It requires sufficient understanding of the targeted bug's root cause and code segment, as well as elaborated engineering effort to create a test failure that captures the buggy behaviors\Space{ with existing testing utilities}, all of which could not be immediately carried out as soon as the bug report is filed.
As a result, bug report sources often prioritize to describing the buggy behaviors over developing a \brt.
Manual effort of developing \brt{}s can also substantially delay fix implementation\Space{, especially when dealing with complex or subtle bugs\samcheng{we didnt show our agent generates brt for complex bug}}.

\subsection{Automated \brt Generation}
\label{sec:movtivation:need}

The importance of {\brt}s in software debugging (\S\ref{sec:movtivation:def}) and the challenge in \brt development (\S\ref{sec:movtivation:available}) motivate the need for automated \brt generation.
Automated \brt generation enables developing \brt{}s for filed bug reports \textit{at scale}, which is especially important in an industrial environment where a large number of bug reports are handled per day.
The generated \brt{}s can then be used by \autopr systems and human developers for bug reproduction, root cause analysis, fix development, and fix validation.
We list below the benefits of automated \brt generation from our experience.

\subsubsection{Improving \autopr Effectiveness}
A prior study on \passerine~\cite{rondon2025passerine}, an \autopr system at \google, suggests that the availability of \brt{}s could aid its fix generation process\Space{for machine-reported bugs}. 
Bug reports of machine-reported bugs often contain \brt information\Space{ (in the form of executable test command)} and textual description of the bug.
\passerine frequently runs the provided \brt and leverages the fault localization information in the \brt while generating fixes for these bugs, resulting in a higher fix rate than on human-reported bugs that have no \brt information.
As our experiment on human-reported bugs will show, generated \brt{}s can also substantially improve \autopr effectiveness by providing valuable context of human-reported bugs and  means to evaluate the generated fixes (\S\ref{sec:results}).

\subsubsection{Enhancing Fix Validation}
Generated \brt{}s can help\Space{ as a reliable mechanism to} validate the effectiveness of proposed fixes, increasing confidence that the bug is truly resolved and preventing regressions~\cite{mundler2024swt,kang2023large}.

\subsubsection{Accelerating Manual Bug Resolution}
Generated \brt{}s provide developers with a readily-usable tool for understanding and diagnosing bugs\Comment{\pmr{``Fixing'' doesn't seem relevant here?}\mt{replaced with "diagnosing", which should fit this paragraph better?}}, which can help reduce their manual reproduction and investigation time~\cite{beller2018dichotomy}.
Automated \brt generation is especially helpful when it transforms a production crash or failure into a reproducible unit test~\cite{soltani2018single,nayrolles2015jcharming}.
Developers can then use familiar debugging tools\Space{ and techniques} to step through the test execution, inspect variables, and isolate the root cause more efficiently and conveniently, compared to directly debugging in a complex production environment or from a limited bug report.

\section{Related Work}
\label{sec:relatedwork}

We first introduce related work in \llm for test generation. 
Next, we discuss recent automated \brt generation techniques, specifically \libro and \sweagentplus from \swtbench. 
We then highlight the similarities and differences between these works and this paper.

\subsection{\llm for Test Generation}
\label{sec:relatedwork:llmtestgen}

The application of \llm{}s to automated test generation has garnered significant attention recently, showcasing the ability of \llm{}s to support developers during testing activities. 
\citet{yuan2023no} employs a conversational approach, iteratively generating unit tests through interaction with \chatgpt. 
This method leverages the conversational abilities of \llm{}s to refine test cases based on ongoing feedback. 
Similarly, \citet{xie2023chatunitest} adopts a feedback-based generation process with \chatgpt, focusing on iterative improvements through continuous interaction.
\citet{lemieux2023codamosa} takes a hybrid approach that combines evolutionary search with the code understanding capabilities of \codex~\cite{chen2021evaluating} to overcome the prior limitation in generating tests that improve coverage.  
\citet{schafer2023empirical} incorporates API documentation alongside \gptthreepfive to generate unit tests, highlighting the importance of providing additional context to \llm for improved test generation.
Further, \citet{li2023nuances} demonstrated that guiding \chatgpt with subtle differences between fixed and buggy codebases can effectively generate failure-inducing tests. 
\Space{Several studies\samcheng{I dont see more than one}}Prior work has empirically evaluated the effectiveness of \llm{}s in unit test generation, e.g., \citet{siddiq2023exploring} investigated and provided insights of \gptthreepfive and \codex on their test generation capabilities. 

The aforementioned studies collectively demonstrate the potential of \llm for automated test generation, showcasing various strategies for prompting, interaction, and context integration.
While these studies offer valuable insights into the application of \llm for test generation, they predominantly study the generation of generic tests, primarily unit tests, whereas we focus on the specific task of generating \brt{}s from bug reports in a production environment.

\subsection{\llm for \brt Generation}
\label{sec:relatedwork:brt}

\subsubsection{\libro}
\label{sec:relatedwork:brt:libro}

\libro~\cite{kang2023large,kang2024evaluating} was one of the first approaches to use \llm{}s for general bug reproduction. 
It frames the task of \brt generation as a few-shot code generation problem, where an \llm is prompted with a few $\langle$bug report, test$\rangle$ examples before being asked to generate a test for the targeting bug report. 
The key assumption of \libro is that \llm{}s, due to their extensive pre-training on vast amounts of code and natural language data, can understand the relationship between a bug description and the test code required to reproduce it.
\libro has the following components~\cite{kang2023large}:
\begin{enumerate}
    \item \textit{Prompt Engineering.} Given a bug report, construct the input prompt with the bug report title and description, and a few $\langle$bug report, test$\rangle$ examples. 
    \item \textit{Querying an \llm.} Query an \llm multiple times with the same input prompt and non-zero temperature to allow variation in the \llm's generated test code each time.
    \item \textit{Test Postprocessing.} Find a test file $F$ that is the most textually similar to each generated test $t$, inject $t$ into $F$ and resolve any new dependencies. \libro considers $t$ a candidate \brt if it compiles and fails on the buggy code.
    \item \textit{Selection and Ranking.} Candidate \brt{}s are clustered and de-duplicated by their failure traces and lines of code. \libro presents one test per cluster in a round-robin fashion to developers, prioritizing tests from the larger clusters. 
\end{enumerate}

\libro shows promising results on \defectsfj 2.0, a benchmark consisting of Java projects, demonstrating the ability of \llm{}s to understand bug descriptions and generate corresponding \brt{}s.\Space{ However, the approach also relied heavily on the quality of the retrieved context and the few-shot examples provided in the prompt, indicating room for improvement in these areas.\samcheng{this is incorrect}}
\citet{ahmed2024tdd} proposed an approach on top of \libro, composed of three \llm calls to find a test file, get function names from bug report, and then generate tests.
Our work, developed independently and concurrently to \citet{ahmed2024tdd}, also found that selecting test file(s) as input to \brt generation is more practical (\S\ref{sec:approach:libro}).

\subsubsection{\sweagentplus}
\label{sec:relatedwork:brt:swtbench}

\Space{\sweagentplus is the best-performing approach on \swtbench~}\citet{mundler2024swt} recently proposed \swtbench, a benchmark built on top of \swebench specifically for \brt generation in Python.
With \swtbench, \citet{mundler2024swt} evaluated \libro. 
They also\Space{ adapted and} evaluated \autopr agents that can interact with the codebase, i.e., \aider~\cite{aider}, \autocoderover~\cite{zhang2024autocoderover}, and \sweagent~\cite{yang2024swe}---these agents were adapted to the \brt generation task via changes in their system and instruction prompts.
On \swtbench, the adapted \sweagent and its proposed variant \sweagentplus perform the best, successfully generating \failtopass tests on 15.9\% and 18.5\% of the instances, respectively. 
They are also the most related to our work.

\sweagent provides an \llm with direct access to augmented command-line tools for searching, viewing, and editing files. 
The agent operates within a limited shell environment and, beyond initial instructions, offers minimal guardrails or structure to its \llm's actions.
Note that one of the tips to \sweagent in its instructions is to start by trying to replicate the bug~\cite{yang2024swe}.
\sweagentplus is the \sweagent variant proposed in \citet{mundler2024swt}, in additional to changes in the system and instruction prompts, it is also explicitly instructed to execute the generated tests before finishing.\Space{Importantly, the instructions do not provide information on how to run the tests. This simple addition was shown to increase the success rate from 15.9\% to 18.5\% in the \swtbench evaluation.}

\subsection{Similarities and Differences with Our Work}
\label{sec:relatedwork:compare}

Like \libro and \swtbench, we focus on \llm for \brt generation.
We adapted and evaluated \libro in our context (\S\ref{sec:approach:libro}).
We also developed an  agent-based approach for automated \brt generation (\S\ref{sec:approach:agent}).
While our approach is developed independently and concurrently to \sweagentplus, they both share the same high-level philosophy in its use of an \llm agent interacting with a codebase via a set of commands to generate \brt{}s.

Our work has the following key differences from prior work:

\myparagraph{Industrial Context}
\libro and \swtbench focus on a set of open-source Java and Python projects that are relatively well-studied~\cite{yang2024swe,Ernst2014Defects4J}, while we focus on \brt generation for production bugs within \google's internal ecosystem~\cite{rondon2025passerine}.
These production bugs are recent, from diverse projects, and span multiple languages\Space{: Java, C++, Go, Python, Kotlin, Dart, and TypeScript~\cite{rondon2025passerine}} (\S\ref{sec:empiricalstudy:datasetandconfig:dataset}).

Developing effective \brt generation for \google's internal environment also poses unique challenges and opportunities, such as dealing with a massive proprietary codebase\Space{ and leveraging}, internal-specific tooling and infrastructure.
For example, when developing our \brt generation agent, we use an \llm fine-tuned on \google's code for the agent's code editing command---this design addresses the confidential nature of \google's code and substantially improves the agent's performance and the effectiveness of its generated \brt{}s.

\myparagraph{Usefulness of the Generated \brt{}s}
In addition to studying the effectiveness of standalone \llm or \llm agent for automated \brt generation, we investigate the practical impact of the generated \brt{}s on an industrial-scale \autopr system---a dimension not explored in detail by \libro or \swtbench.

Specifically, we study the generated \brt{}s on\Space{ two use-cases}: (1) their effectiveness on improving the performance of \autopr system; (2) their effectiveness on selecting the most plausible \autopr-generated fixes.
Our study adds another layer of practical application not explicitly addressed in the related work in \llm-based \brt generation.

\section{Approaches}
\label{sec:approach}

\begin{figure*}[htbp] %
  \centering

  \begin{subfigure}[b]{\textwidth}
    \centering
    \includegraphics[width=0.8\textwidth]{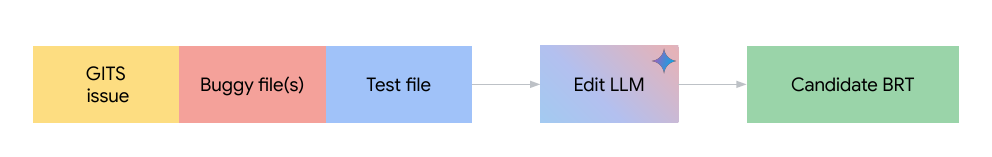}
    \caption{\libro Adaptation: The input comprises the bug report (\gits issue) accompanied by the list of buggy files and the relevant test file. \libro  prompts the \codeeditingllm to generates a candidate \brt\Space{ for the given issue}.}
    \label{fig:libro}
  \end{subfigure}

  \vspace{1em} %

  \begin{subfigure}[b]{\textwidth}
    \centering
    \includegraphics[width=0.8\textwidth]{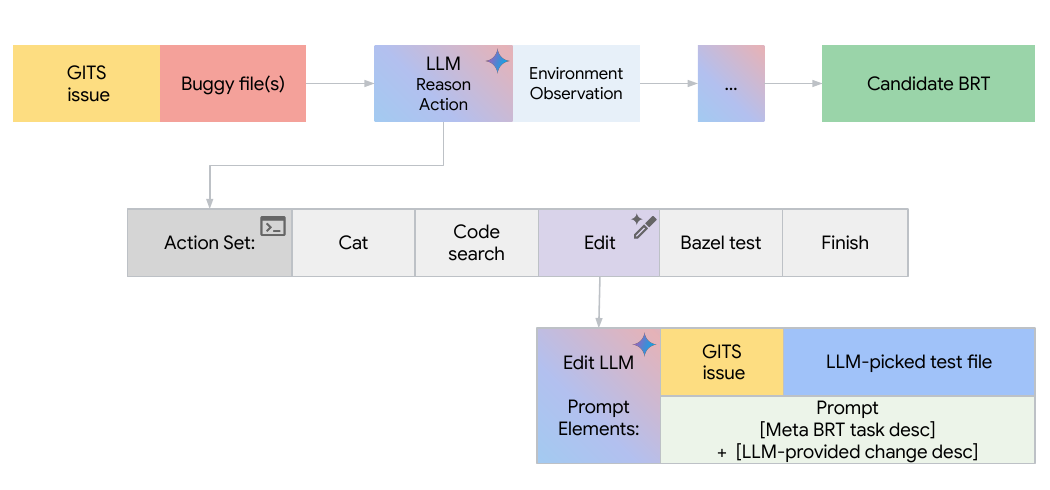}
    \caption{BRT Agent (Ours): Given the bug report (GITS issue) and buggy file content as initial input, our \llm-agent-based approach leverages a set of actions, including code editing via the \codeeditingllm, to generate a candidate BRT.}
    \label{fig:brtagent}
  \end{subfigure}

  \caption{Bug Reproduction Test (BRT) generation techniques explored in our work.}
  \label{fig:main_figure}
\end{figure*}

In this section, we first describe our adaptation of \libro to Google's internal development environment.
We then detail the design and implementation of our agent-based approach for \brt generation, which we will refer to as \textbf{\tool}.
Figure~\ref{fig:main_figure} illustrates both our adaptation of \libro (\ref{fig:libro}) and \tool (\ref{fig:brtagent}).\Comment{This section details the design and implementation of our approach for automated BRT generation within Google's internal infrastructure. 
We first describe our adaptation of the baseline technique, LIBRO, to function effectively in this unique environment. 
Subsequently, we introduce our enhanced BRT agent, designed to overcome the limitations of existing methods by leveraging a fine-tuned LLM for code editing.}

\subsection{Adaptation of \libro}
\label{sec:approach:libro}

\libro was originally built on open-source projects. 
We make changes to \libro's \llm, input data, and prompt to make it compatible with \google's internal development environment.

\myparagraph{\llm}
We replaced the open-source \llm that \libro uses with a Gemini model fine-tuned on \google's internal code. 
This \llm is exposed to the specific coding styles, libraries, and patterns prevalent within \google's codebase, and can improve \libro's ability to generate relevant and syntactically correct code. 
Note that we also use this fine-tuned \gemini model as the \codeeditingllm for our agent-based \brt generation approach (\S\ref{sec:approach:agent}). 

\myparagraph{Input Data}
\libro finds a relevant test file to inject a \brt into and resolves dependencies after the \brt has been generated (\S\ref{sec:relatedwork:brt:libro}). 
We find this workflow error-prone and impractical within \google's complex and interconnected codebase. 
Thus, we adapt \libro to accept a pre-identified test file as input, which can provide useful test code context (e.g., libraries and attributes used by existing tests) to the \llm, to generate a more robust and relevant \brt.\Comment{substantially simplifies the generation process and ensures that the generated \brt is compatible with existing testing infrastructure.} 
In practice, the input test file(s) will be identified as the test file most related to the buggy file(s), e.g., via code search or coverage analysis.

\myparagraph{Prompt}
We carefully crafted the input prompt for the internal \llm used by our adapted \libro through small-scale experiments on different prompt structures, paraphrases, and input.
The final prompt includes the bug report description, buggy file content, and a pre-identified test file.
\Comment{commented out because original \libro has this too. \textbf{Multiple Sample Generation}: We modified LIBRO to enable it to generate multiple sample BRTs by calling the LLM multiple times, using a non-zero temperature. This allows us to explore the diversity of generated tests and potentially identify more effective BRTs.}

\subsection{Our Approach: \tool}
\label{sec:approach:agent}

Recent research on \llm agents for \autopr have shown promising results~\cite{zhang2024autocoderover,yang2024swe,chen2024coder,hou2024large,rondon2025passerine}\Space{, and is actively investigated in industrial settings~\cite{rondon2025passerine,OTHER_STUDIES}}, while \llm agents for \brt generation remained rather under-explored.
Moreover,\Space{ in our experience,} we find that state-of-the-art technique using standalone \llm, i.e., \libro, faces performance challenges\Space{ on large-scale, proprietary codebase} even after adaptation (\S\ref{sec:results}).
Thusly motivated, we develop an agent-based approach for \brt generation: \tool.\Space{\tool iteratively improves the generated \brt via reasoning and performing actions according to its observed environments \react~\cite{yao2022react,yang2024swe}.}
We now describe \tool's workflow and components.

\subsubsection{Initialization}

\tool receives a bug report and identified buggy file content\Comment{\sam{no test file}, and the path to the chosen test file} as initial input.\Comment{we should mention it in experiment setup instead \textit{In our evaluation, an oracle provides the buggy code files and test file path. In practice, a user may provide this information.}} These files can be provided by users or upstream fault localization services~\cite{wong2016survey,kang2023preliminary,yang2024large,qin2024agentfl}.

\subsubsection{Reasoning}
\label{sec:approach:agent:reasoning}
A \gemini model serves as the primary reasoning \llm, responsible for understanding the bug report, planning overall execution of the \brt generation task, and making decisions about which action to take in each \react step~\cite{yao2022react}.
At each step, it analyzes the current agent state, which includes the bug report, the displayed code context, and the history of past actions and observations, to eventually determine the next action.

\subsubsection{Action Selection and Execution}
\label{sec:approach:agent:action}

\begin{table*}[htbp]
\centering
\begin{adjustbox}{max width=\textwidth}
\begin{tabular}{p{0.17\textwidth} p{0.83\textwidth}}
\toprule
\textbf{Action} & \textbf{Description} \\ \midrule
\CodeIn{cat [path]} & Displays the content of the file at the specified path. This allows the agent to inspect code files. \\ \midrule
\CodeIn{code\_search [text]} & Searches \google's internal code repositories for code snippets matching the given query. This enables the agent to find relevant code examples or identify potential locations of the bug. \\ \midrule
\CodeIn{edit [path] [prompt]} & \Space{A specialized command that t}Triggers the code editing process\Space{, as detailed in step 4 below}. It is designed to prompt an \llm fine-tuned for editing in \google's codebase. To avoid introducing errors to non-test code, we restrict \texttt{edit} to only apply to test files identified based on regex\Space{\CodeIn{*[tT]est\..*}}. \\ \midrule
\CodeIn{bazel test [path]} & Executes the \CodeIn{bazel test} command on a specified target via \google's internal testing framework. This allows agent to run tests and observe the results. We also de-duplicate test output if the test log is too long for \llm input context. \\ \midrule
\CodeIn{finish} & Signals the agent's belief that it has either successfully generated a \brt or that it cannot create a \brt. The agent is restricted from using \CodeIn{finish} before running any tests. \\
\bottomrule
\end{tabular}
\end{adjustbox}
\caption{\tool actions.\label{tab:actions}}
\vspace{-10pt}
\end{table*}

Table~\ref{tab:actions} shows a list of possible actions \tool can select and execute. For action \actioncat, \actioncodesearch, \actiontest, and \actionfinish, the agent executes the action directly.
For action \actionedit, the agent initiates a specialized code editing process that will be handled by a \codeeditingllm. Specifically:

\myparagraph{Change Description}
As part of the \actionedit action invocation, \tool, via its reasoning \llm, generates a natural language description of the desired code change to a test file the agent specifies.
The agent often chooses a test file from results of \actioncodesearch action(s) from prior steps (\S\ref{sec:results:brt:behave}). 
The description captures the intent of the edit; one example description can be: ``Add a test case that asserts the function returns \CodeIn{null} when given an empty input.''

\myparagraph{Prompt Construction}
The action \actionedit constructs a prompt for the \codeeditingllm, which includes (1) the bug report description; (2) the content of the specified test file; and (3) a clear task description composed of two parts: a \textit{meta} task description fixed to ``Write a bug reproduction test for the following bug report.'', and a \textit{step-specific} change description, which describes the desired change in natural language, generated by the reasoning \llm for this \actionedit invocation.

\myparagraph{Code Generation}
The \codeeditingllm receives the prompt and generates a code patch that implements the described change. 
The generated patch will then be applied to the specified test file.

The \codeeditingllm, given it is fine-tuned on \google's internal codebase, offers the following key advantages. 
First, by being exposed to a vast amount of internal code, the \codeeditingllm possesses a deep understanding of the nuances and context specific to the production environment, making it more effective in understanding and thus reproducing a bug.
Second, the \codeeditingllm can generate code that seamlessly integrates with \google's production infrastructure while adhering internal coding standards, ensuring the generated test to be functional and align with established best practices and conventions.
Employing the \codeeditingllm can thus improve the agent's ability to generate syntactically- and semantically-correct \brt{}s.

Moreover, this design of combining generic \llm (which demonstrated reasoning proficiency) with fine-tuned \codeeditingllm could be generally applicable and efficient to cases where the software engineering agent is deployed to a codebase with new APIs and conventions outside the generic \llm's training distribution.

Note that the \codeeditingllm's training data excludes all bugs, code changes, and \brt{}s used in our empirical evaluation (\S\ref{sec:empiricalstudy})---its training data cutoff predates the reporting of all bugs analyzed in this study, preventing any potential data leakage.

\subsubsection{Observation}
\label{sec:approach:agent:observe}
\tool observes the result of its previous executed action. 
Action \actiontest produces one of these test results, accompanied with test logs: 
    \begin{itemize}
        \item \textbf{Pass}: All specified tests executed successfully and passed.
        \item \textbf{Fail}: All specified tests executed but at least one test failed, indicating a potential bug reproduction.
        \item \textbf{Broken/Error}: The specified tests could not be executed successfully due to compilation errors, runtime errors before test execution, or other issues in the setup. This state often signifies that the code edits require further refinement.
    \end{itemize}

\subsubsection{Iteration}
\label{sec:approach:agent:iterate}
The agent repetitively attempts \S\ref{sec:approach:agent:reasoning}-\ref{sec:approach:agent:observe} until it finishes (see Table~\ref{tab:actions}) or exhausts the total number of allowed steps.

\subsubsection{Termination}
\label{sec:approach:agent:terminate}
\tool terminates gracefully by invoking the action \actionfinish, or reaching the total step limit. 
The final output will be the generated test code, if any, within a test file.

\section{Empirical Study Design}
\label{sec:empiricalstudy}

This section outlines the design of our empirical study\Space{ on the effectiveness of our \brt generation techniques}. 
We describe the evaluation dataset, experiment configurations, research questions (RQs)\Space{ guiding our investigation}, and evaluation metrics for each RQ\Space{ employed to assess the performance of our approaches}.

\subsection{Dataset and Experiment Configurations}
\label{sec:empiricalstudy:datasetandconfig}

\subsubsection{Dataset}
\label{sec:empiricalstudy:datasetandconfig:dataset}
Our evaluation is conducted on a dataset of\Space{ approximately} 80\footnote{Concurrent work~\cite{rondon2025passerine} reports 78 bugs in the same set due to recent, unrelated infrastructure changes rendering two bugs' tests no longer executable; this work was performed before the change.} production bugs in\Space{collected from} \google's internal issue tracking system (\gits).
All the bugs were reported by human developers, and were fixed by human developers from corresponding teams.
This dataset was constructed via automated extraction and filtering phases as well as manual curation\Space{, in previous research conducted at Google}~\cite{rondon2025passerine}. 
The bugs are recent (since June 2024), from a wide array of \google projects, and span seven programming languages: Java, C++, Go, Python, Kotlin, Dart, and TypeScript.\Space{Specifically, we focus on\Space{ the subset of} bugs reported by human developers, and exclude those reported by automated tools from the dataset.}
In the manual curation process, each bug and its corresponding fix were carefully reviewed to ensure that the fix genuinely addresses the underlying issue reported in the bug report, and not merely a workaround or masking the bug. 

Each bug sample in the evaluation dataset has:

\begin{itemize}
    \item \textbf{\gits Issue}: A bug report with a title and a description\Space{ of the issue}, reported by a human developer.
    \item \textbf{Ground Truth Fix}: A code change (CL) that resolves the reported bug. 
    This fix includes an ``oracle'' \brt---a manually written test that reproduces the bug and validates the fix~\cite{frommgen2024resolving}. 
    \Space{Importantly, due to the manual curation, these fixes are}Each fix has been manually verified to address the root cause of the corresponding issue as identified in the bug report~\cite{rondon2025passerine}.
    \item \textbf{Ground Truth Buggy Files}: The files modified by the ground truth fix, representing the buggy version of the code.
    \item \textbf{Ground Truth Test File}: The test file modified by the ground truth fix to include the oracle \brt.
\end{itemize}

\subsubsection{Experiment Configurations}
We evaluate the adapted \libro (\S\ref{sec:approach:libro}) and our \tool (\S\ref{sec:approach:agent}) for \brt generation.

We provide the \gits issue title and description, ground truth buggy files, and the ground truth test file to the adapted \libro as input. 
We provide only the \gits issue title and description, along with the ground truth buggy files, to \tool as input.

\libro uses only one \llm, while \tool uses a reasoning \llm and a \codeeditingllm (\S\ref{sec:approach}).
We use the same \gemini model fine-tuned on \google's internal code as \libro's \llm and \tool's \codeeditingllm, and use a publicly available \gemini as \tool's reasoning \llm. 
For each $\langle$technique, bug$\rangle$, we perform multiple experiment runs to account for the stochastic nature of \llm{}s. 
For \libro, we set a temperature of 0.7 and a top P of 0.95 for the \llm, and perform 50 runs per bug~\cite{kang2023large}. 
For \tool, we set a temperature of 0.2 and a top P of 0.95 for both the reasoning and the \codeeditingllm, and perform 20 runs per bug~\cite{rondon2025passerine}. 
We set the total number of steps the \tool can take per run to 25. 

We provide three synthetic \brt generation examples in the system prompt of \tool to provide additional context of the \brt generation task to \gemini\Space{: an example where a \brt generation is done with one \actionedit{}-\actiontest iteration, and two examples where multiple \actionedit{}-\actiontest iterations were performed before the \brt is generated}.  
These examples demonstrate the desired agent behaviors in a \react format, showcasing trajectories of successful and unsuccessful \brt generation attempts, and were unchanged across all experiments. 
We did not provide these examples to the adapted \libro because it is not an agent-based approach. 
We need not provide code generation examples to \libro because it already uses a \codeeditingllm trained for \google internal code generation tasks, including test generation.\Comment{We drop the selection and ranking component of \libro because we evaluate all of its generated \brt{}s.}

\subsection{Research Questions}
\label{sec:empiricalstudy:rq}
Our study aims to answer the following research questions:

\begin{itemize}
    \item \textbf{RQ1: \rqonetitle}: How effective is our \tool in generating \brt{}s for \google's internal human-reported bugs?
    \item \textbf{RQ2: \rqtwotitle}: How effective are the generated \brt{}s for improving the bug-fixing performance of an \autopr system?
    \item \textbf{RQ3: \rqthreetitle}: How effective are the generated \brt{}s for selecting plausible fixes generated by an \autopr system?
\end{itemize}

\subsubsection{RQ1: \rqonetitle}
\label{sec:empiricalstudy:rq1}
A high success rate in generating \brt{}s in an industrial context is crucial, because it provides developers with a valuable tool for understanding, debugging, and ultimately fixing software defects, thereby\Space{ significantly} reducing the\Space{ manual} effort\Space{ typically} required in the debugging process and accelerating the overall software development lifecycle. 
Successfully developing \brt generation techniques for an industrial environment like \google's would demonstrate their practical applicability and potential for broader impact beyond\Space{ the realm of} open-source projects.

\myparagraph{Setup}
We run \libro and \tool with settings as described in \S\ref{sec:empiricalstudy:datasetandconfig}.
We compute their \brt generation effectiveness with the same metrics as recent studies~\cite{mundler2024swt,kang2023large}:\Space{\samcheng{lets not mention inspection here just like agent behavior analysis}}

\begin{itemize}
    \item \textbf{Candidate \brt{}s}: Percentage of bugs for which at least one test is generated that fails on the buggy version of the code (\failtoany). 
    These are considered candidate \brt{}s because failing on the buggy version is a necessary, but not sufficient, condition for a test to be a true \brt.
    \item \textbf{Plausible \brt{}s}: Percentage of bugs for which at least one test is generated that fails on the buggy version and passes on the fixed version of the code (\failtopass). 
    These are considered plausible \brt{}s because they satisfy both necessary conditions for reproducing a bug and validating the fix.
    \item \textbf{Candidate to Plausible \brt Rate}: Percentage of bugs with candidate \brt{}s that are confirmed to be as plausible \brt{}s when applied to the ground truth fix. 
    This metric provides insight into the precision of the \brt generation technique.
\end{itemize}

\subsubsection{RQ2: \rqtwotitle}
\label{sec:empiricalstudy:rq2}

This research question investigates the practical impact of the automatically generated \brt{}s on fix generation. 
The hypothesis is that \brt{}s provide additional context to better understand the buggy behaviors, which can improve bug-fixing efficiency (\S\ref{sec:motivation}).\Space{the hypothesis should be kept because we use plausible \brt not candidate \brt.}

Specifically, we want to understand how providing \brt{}s to \passerine~\cite{rondon2025passerine}, an industrial-scale, \llm-agent-based \autopr system designed to work with \google's internal codebase and framework, can improve its ability in generating plausible fixes.

\myparagraph{Setup}
For each bug in which \tool did generate plausible \brt{}(s), we provide a generated plausible \brt{} to \passerine by (1) applying the generated \brt patch to the buggy codebase version where \passerine will perform the bug-fixing activities, and (2) mentioning a \brt (with test path provided) is available in the \gits issue.
We let \passerine decide whether to use the provided \brt\Space{ during its bug-fixing run}.
Note that the plausibility of a \brt is determined with a fix (\S\ref{sec:empiricalstudy:rq1}), results from this research question thus represent an upper bound on the impact of generated \brt{}s on \autopr\Space{, independent of the \brt generation strategy\samcheng{we use brts from \tool}}.

To properly sample a generated plausible \brt for each bug, and allow such sampling procedure to be applicable in practice where a fix would not be available prior to the sampling, we use \gemini as \llm-as-a-Judge~\cite{zheng2023judging}.
We prioritize selecting a plausible \brt{} that (1) has a test failure judged to be caused by the bug, (2) was generated from a gracefully terminated \tool run, and (3) its last execution in the \tool run produced a test failure. 
We sample randomly if multiple or no candidate meet these criteria.

We compare \passerine's performance with versus without providing a generated \brt, on the subset of 23 bugs where \tool generated plausible \brt{}s. 
We run \passerine with \gemini on temperature 0.2 and top P 0.95 for 20 runs per bug~\cite{rondon2025passerine}. 
We determine the plausibility of a generated fix by running the oracle \brt on it.

We use these metrics to measure \passerine's performance:
\label{sec:empiricalstudy:rq3}
\begin{itemize}
    \item \textbf{Number of Bugs with Plausible Fixes}: Number of bugs for which \passerine generates at least one plausible fix (a fix that passes the oracle \brt which \passerine cannot access).\Space{ We compare this number between runs with and without the generated \brt provided as input.}
    \item \textbf{Steps to Plausible Fix}: The\Space{ average} number of steps taken by \passerine to generate a plausible fix\Space{, comparing runs with and without the generated \brt}.
    \item \textbf{Plausibility Given \brt Usage}: The probability that a generated fix is plausible given that the generated \brt was used by \passerine in the bug-fixing run.
\end{itemize}

\subsubsection{RQ3: \rqthreetitle}

This research question explores the utility of generated \brt{}s as a mechanism for selecting plausible fixes from a pool of possible fixes generated by an \autopr system. 
\llm-based \autopr systems can generate a large number of potential fixes per bug~\cite{jiang2023impact,fan2023automated,xia2023automated,jimenez2023swe,rondon2025passerine}, efficiently identifying a correct fix from all potential fixes is a crucial challenge when \textit{no existing test} had failed because of the bug\Space{, i.e., no existing test is a \brt}. 
We thus examine whether\Space{ the suite of} generated \brt{}s can be used to effectively discriminate between correct and incorrect fixes, and to rank fixes based on their likelihood of being correct. 

\newcommand{\passtopass}{$P \rightarrow P$\xspace}

We introduce \emph{\enpassratefull (\enpassrate)}, defined as the percentage of candidate \brt{}s that pass when executed against a given fix. 
The intuition is to use the generated tests (i.e., candidate \brt{}s with \failtoany behavior) as a ``synthetic test suite'' to assess the effectiveness for all potential fixes, and leverage the pass/fail test results to inform fix selection.
We omit Pass-to-Pass tests~\cite{mundler2024swt} because they do not fail on the bug.
\Comment{This approach aims to balance between checking the validity of fix with only one~\cite{kang2023large} or all~\cite{mundler2024swt} generated \brt{}s.\samcheng{not sure if this has been mentioned, lets not draw attention}}
If successful, this approach would be a useful tool for improving the accuracy of \autopr systems and reducing the manual effort required to validate their outputs.

\myparagraph{Setup} 
Different from \S\ref{sec:empiricalstudy:rq2}, in this research question, \passerine is not provided a \brt for its bug-fixing run---we only evaluate the usefulness of the generated \brt{}s for fix selection,\Space{ in isolation} after \passerine has generated fixes.
We use standard \irfull metrics to measure the efficacy of \enpassrate in selecting plausible fixes: 

\begin{itemize}
    \item \textbf{Precision, Recall, F1-score, and Mean Reciprocal Rank (MRR)}.
    We define a true positive as a selected fix that passes the oracle \brt, a false positive as a selected fix that fails the oracle \brt, and a false negative as an unselected fix that passes the oracle \brt. 
\end{itemize}

\Space{We analyze how these metrics vary with different values of an integer or a float threshold. }We compute values of these metrics for \enpassrate-based fix selection under two settings: (1) \textit{Top-K Selection}, where we select the top-$K$ fixes with the highest \enpassrate, and (2) \textit{Threshold-Based Selection}, where we select all fixes with an \enpassrate above a certain threshold.

\section{Empirical Results}
\label{sec:results}
We now present results of our research questions defined in \S\ref{sec:empiricalstudy:rq}:
\begin{itemize}
    \item \textbf{RQ1: \rqonetitle}: How effective is our \tool in generating \brt{}s for \google's internal human-reported bugs?
    \item \textbf{RQ2: \rqtwotitle}: How effective are the generated \brt{}s for improving the bug-fixing performance of an \autopr system?
    \item \textbf{RQ3: \rqthreetitle}: How effective are the generated \brt{}s for selecting plausible fixes generated by an \autopr system?
\end{itemize}

\subsection{\rqonetitle\Space{Bug Reproduction Effectiveness}}
\label{sec:results:brt}

\subsubsection{Overall Effectiveness}
Table \ref{tab:rq1_results} presents the \brt generation effectiveness of the evaluated techniques.  
The adapted \libro achieves a candidate \brt generation rate of 41\%. Its plausible \brt generation rate is 10\%\Space{. This}, which is lower than the 33\% reported on \defectsfj~\cite{kang2023large}, potentially due to the increased complexity of Google's internal bugs~\cite{rondon2025passerine}. 
The main failure mode is represented by \libro generating \brt{}s with build errors, from which it cannot recover.

On the other hand,\Space{ Our} \tool demonstrates superior performance. 
It achieves an 85\% candidate \brt generation rate and a 28\% plausible \brt generation rate, significantly outperforming \libro. 
The higher candidate-to-plausible \brt rate (34\%) indicates that \tool is more effective at generating tests that are likely to be actual \brt{}s.

Table~\ref{tab:rq1_languages} provides a detailed breakdown of plausible \brt{}s by programming languages.
Each value in Table~\ref{tab:rq1_languages} is computed relative to the number of evaluated bugs in a language.
Our \tool generates plausible \brt{}s across 6 languages\Space{: Java, C++, Go, Python, Kotlin, and TypeScript}.
It generates plausible \brt{}s on the same and higher number of bugs than \libro in 3 and 4 languages, respectively.
The ability to generalize across multiple languages is a particularly important characteristic of agent-based systems, as it suggests a broader applicability to diverse codebases and development environments~\cite{yang2024swebenchmultimodalaisystems}.\Space{\samcheng{i am not sure what this sentence means but fine for me to keep}}

Our results show that, \tool, by generating \brt in an agentic fashion on top of a \codeeditingllm specialized on \google's internal code, is significantly more effective in the industrial setting.\Space{\samcheng{add that \tool did not find the right test in 10\% of the cases}---\samcheng{this one is hard to argue, not mention}}

\begin{table}[t!]
\centering
\begin{tabular}{lcc}
\toprule
Metric & LIBRO & \tool (Ours) \\
\midrule
Candidate BRTs & 41\% & 85\% \\
Plausible BRTs & 10\% & 28\% \\
Candidate to Plausible BRTs & 24\% & 34\% \\
\bottomrule
\end{tabular}
\caption{\brt generation effectiveness.}
\label{tab:rq1_results}
\end{table}

\begin{table}[t!]
\centering
\begin{tabular}{lcc}
\toprule
Languages & \libro & \tool (Ours) \\
\midrule
Java & 8\% & 28\% \\
C++ & 0\% & 16\% \\
Go & 17\% & 17\% \\
Python & 18\% & 45\% \\
Kotlin & 10\% & 50\% \\
Dart & 0\% & 0\% \\
TypeScript & 100\% & 100\% \\
\bottomrule
\end{tabular}
\caption{Plausible \brt{}s by languages. Each percentage value is relative to the number of bugs in a language.}
\label{tab:rq1_languages}
\end{table}

\myparagraph{Manual Inspection}

To further investigate \tool's generation quality, we manually inspect the generated patches of all plausible \brt{}s, by comparing them against the oracle \brt{}s.
Specifically, two authors inspect the \brt{}s and a third author resolves any disagreement.\Space{michele: we should give a bit of information about how this manual analysis was conducted: 2 authors analyzed the BRTs + 1 author resolved the conflicts on 5 cases.}
We consider a plausible generated \brt patch to be \textit{valid} if (1) it is identical or semantically equivalent to the oracle \brt patch, and (2) it did not modify any existing test.

Through inspection, we find that 86\% of the plausible generated \brt patches are valid, or 25\% of the bugs have at least one valid \brt patch.
The percentage of plausible \brt patches that are identical and semantically equivalent to the oracle \brt{}s are 19\% and 48\%, respectively.
Meanwhile, 16\% of the plausible \brt patches, despite being valid, add irrelevant assertion(s) or new test(s) that pass on the bug.
We reason these irrelevant code segments are by-products of the agent's attempts into iteratively improving its previous edits~\cite{jimenez2023swe,rondon2025passerine} before producing a test failure.
We also find that 11\% of the plausible \brt patches to be invalid because they have modified existing test(s).
Our inspection results imply that the majority of generated plausible \brt patches are readily-usable, and future work can further improve \tool by systematically resolving its suboptimal behavior of changing existing tests into valid \brt{}s.

\subsubsection{Agent Behaviors}
\label{sec:results:brt:behave}
We also study the statistics on agent behaviors, specifically the action distribution and termination reasons.

\begin{figure*}[t!]
    \centering
    \includegraphics[width=0.8\linewidth]{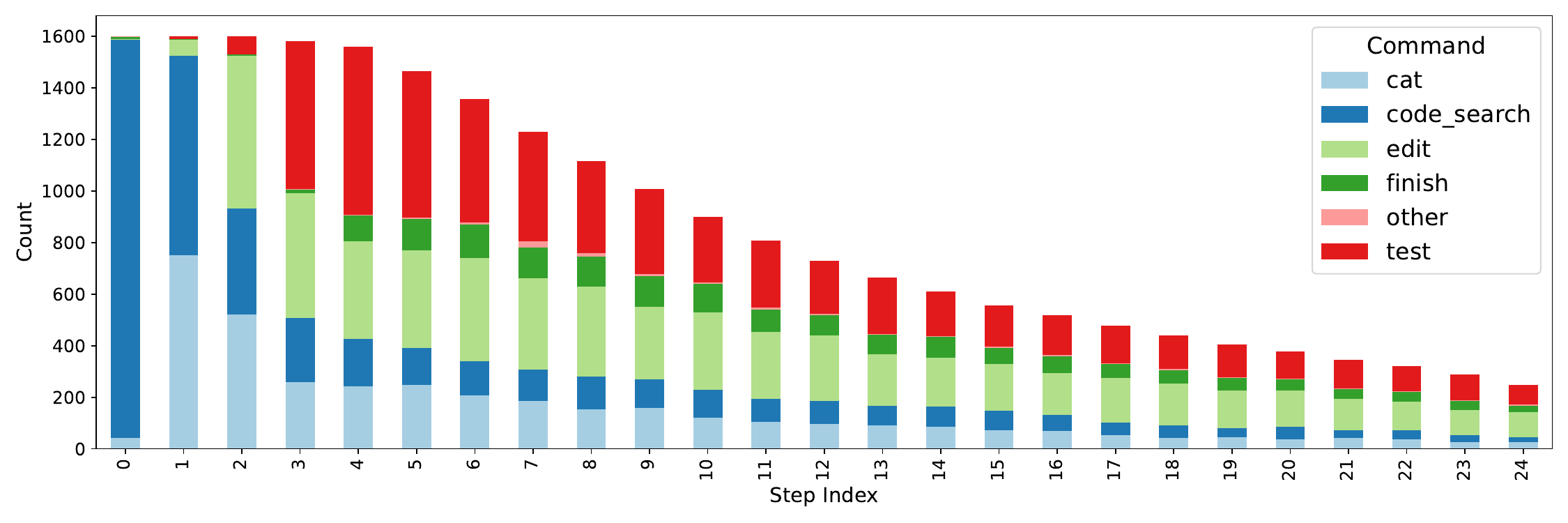} 
    \caption{Action distribution by step for the \tool.\label{fig:action_distribution}}
\end{figure*}

Figure \ref{fig:action_distribution} shows the action distribution of each step for all 1600 (80 * 20) runs of \tool. 
In each run, the agent can take at most 25 steps (\S\ref{sec:empiricalstudy:datasetandconfig}).
The ``other'' category includes actions output by the agent that are not part of the provided action set---in a few cases, the agent hallucinated and output actions, such as \CodeIn{grep} or \CodeIn{find}, that are not listed in its system prompt.
Figure~\ref{fig:action_distribution} also shows \tool frequently starts by finding related files via \actioncodesearch, inspecting their content via \actioncat, and  proceeding to edit and run tests.

\begin{table}[t!]
\centering
\begin{tabular}{lc}
\toprule
Top-5 action bigram & Frequency \\
\midrule
(\CodeIn{edit}, \CodeIn{bazel test}) & 19\% \\
(\CodeIn{cat}, \CodeIn{edit}) & 10\% \\
(\CodeIn{bazel test}, \CodeIn{edit}) & 9\% \\
(\CodeIn{code\_search}, \CodeIn{code\_search}) & 9\% \\
(\CodeIn{code\_search}, \CodeIn{cat}) & 8\% \\
\bottomrule
\end{tabular}
\caption{Top-5 most frequent action bigrams of \tool.}
\label{tab:action_bigrams}
\end{table}

Table \ref{tab:action_bigrams} shows the top-5 most frequent action bigrams from \tool. 
The frequency column indicates the occupancy of a specific bigram among all bigrams. 
The top-3 bigrams mainly represent the agent's core task of \brt generation: editing, analyzing, and running test files. 
The two remaining bigrams mainly represent the agent's file localization and debugging activities.

\begin{table}[t!]
\centering
\begin{tabular}{lcc}
\toprule
Termination reason & \libro & \tool (Ours) \\
\midrule
Requested termination & 83\% & 72\% \\
Steps exhausted & - & 21\% \\
Framework exception & 17\% & 7\% \\
\bottomrule
\end{tabular}
\caption{Termination reasons of techniques.}
\label{tab:termination_reasons}
\end{table}

Table \ref{tab:termination_reasons} shows the category of reasons that the evaluated \brt generation techniques terminated.
\tool\Space{ frequently} terminates gracefully by invoking the \actionfinish action in 72\% of the cases, while exhausting the configured total number of steps (i.e., 25) in 21\% of the cases. 
There are also cases where \tool and \libro throw framework-level exceptions; common reasons for framework-level exceptions are the \llm input exceeding the input context window size, server rate limit exceptions, and arbitrary exceptions thrown from imported libraries outside the framework (\tool only).

\subsection{\rqtwotitle\Space{Impact on Automated Patch Generation}}
\label{sec:results:generation}

We now present results of \passerine's bug-fixing effectiveness with versus without providing generated plausible \brt{}s (\S\ref{sec:empiricalstudy:rq2}).

\begin{figure}[t!]
    \centering
    \includegraphics[width=0.8\columnwidth]{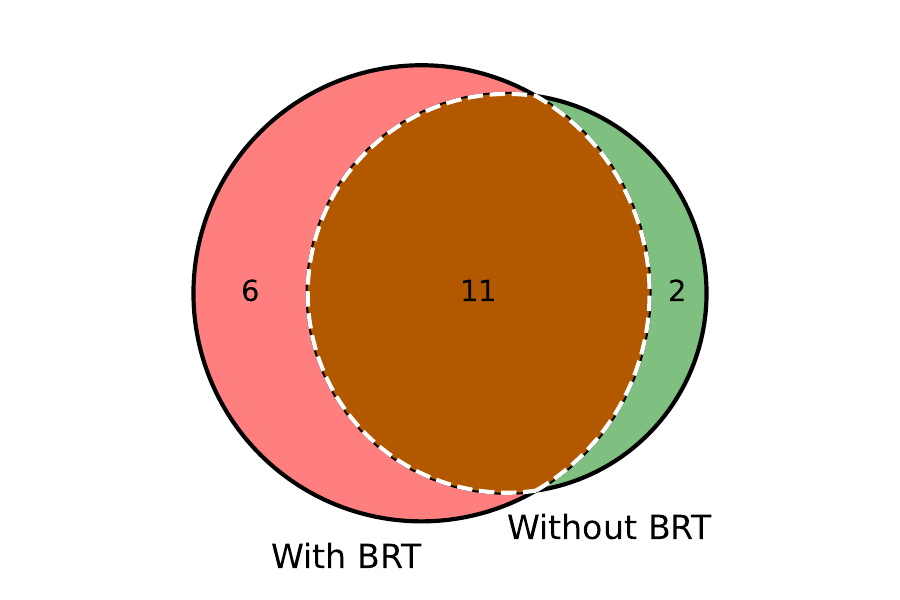}
    \caption{The number of plausible fixes generated by \passerine with (red) and without (green) \brt as input.}
    \label{fig:rq2_plausible_fixes}
\end{figure}

\myparagraph{Plausible Fixes}
Generated \brt{}s from \tool can substantially improve \passerine's performance in generating plausible fixes.
Figure~\ref{fig:rq2_plausible_fixes} shows the sets of bugs resolved by \passerine with versus without providing \brt{}s as initial input. 
\passerine generated plausible fixes in 74\% (17/23) of the bugs when provided with a generated \brt as initial input, compared to 57\% (13/23) when not provided a \brt. 
Providing \brt{}s help \passerine solve 6 new, unique bugs that it would have not solved if without a \brt.

\begin{figure}[t!]
    \centering
    \includegraphics[width=\columnwidth]{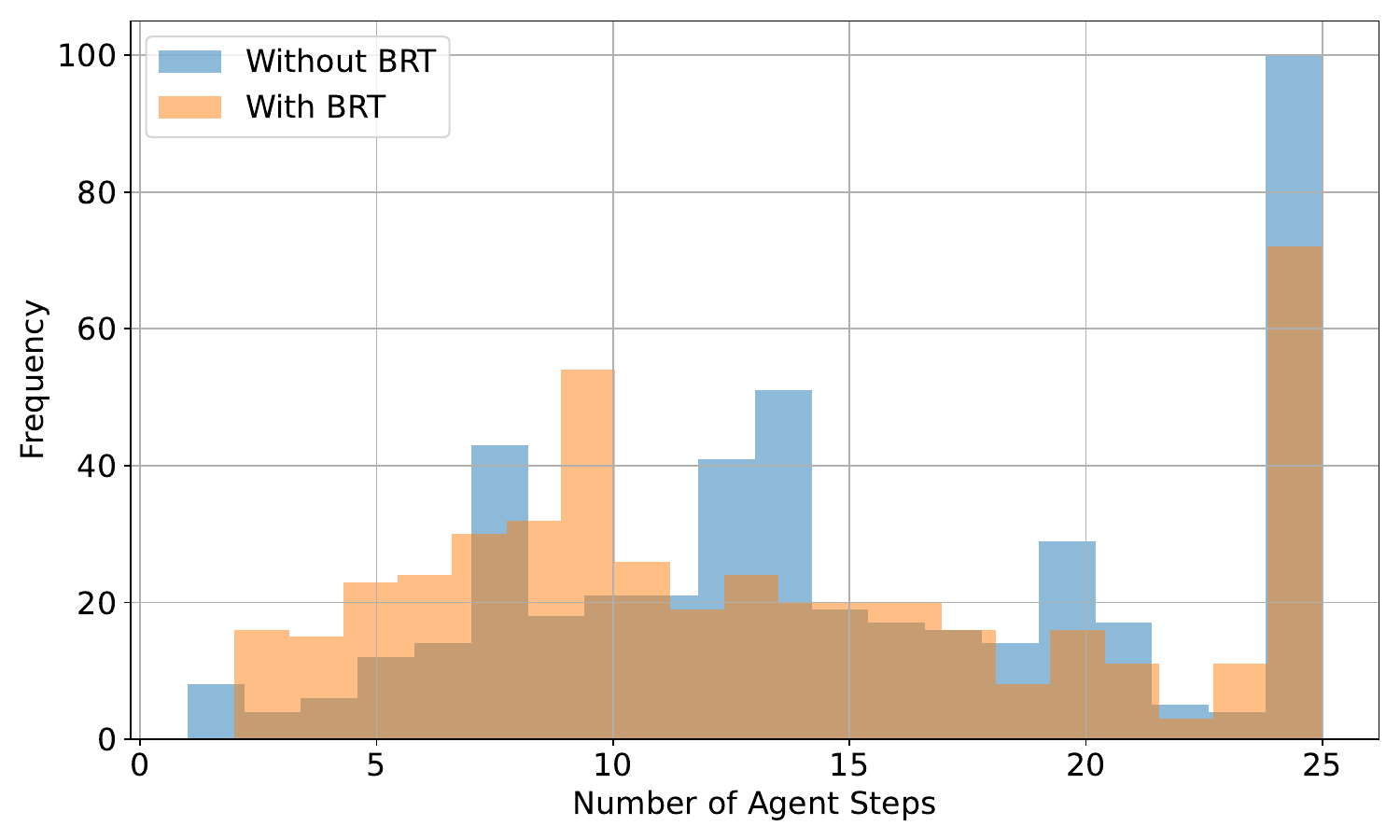}
    \caption{Average number of steps per run for \passerine to generate plausible fix with and without \brt as input.}
    \label{fig:rq2_steps_to_fix}
\end{figure}

\myparagraph{Steps to Plausible Fixes}
\passerine took fewer steps on average to generate plausible fixes when provided with a \brt. 
Figure \ref{fig:rq2_steps_to_fix} depicts the frequency distributions of \passerine's runs completed within $k$ steps for both settings (i.e., with and without \brt), with a maximum of $k=25$ steps. 
The leftward shift from the distribution of ``without \brt'' to the distribution of ``with \brt'' indicates that Passerine resolved bugs with fewer steps when provided a \brt.

\myparagraph{Probability of Plausible Fixes}
The probability of \passerine generating a plausible fix given that the generated \brt was used is 33\%, compared to only 2\% when the \brt was not used. 
This observation implies that a plausible fix is more likely to be generated by the \autopr agent in runs where the \brt is provided and used.

Overall, providing generated \brt{}s to an \autopr system like \passerine could improve its bug-fixing performance, both in terms of the number of bugs fixed and the fix generation efficiency. 
Meanwhile, \autopr agents should also attempt to generate multiple fixes both with and without providing \brt{}s through sampling to cover more inference trajectories and unlock new possibilities.

\subsection{\rqthreetitle\Space{Impact on Automated Patch Selection}}
\label{sec:results:selection}

\subsubsection{Top-K Selection}
\begin{figure}[t!]
    \centering
    \includegraphics[width=\columnwidth]{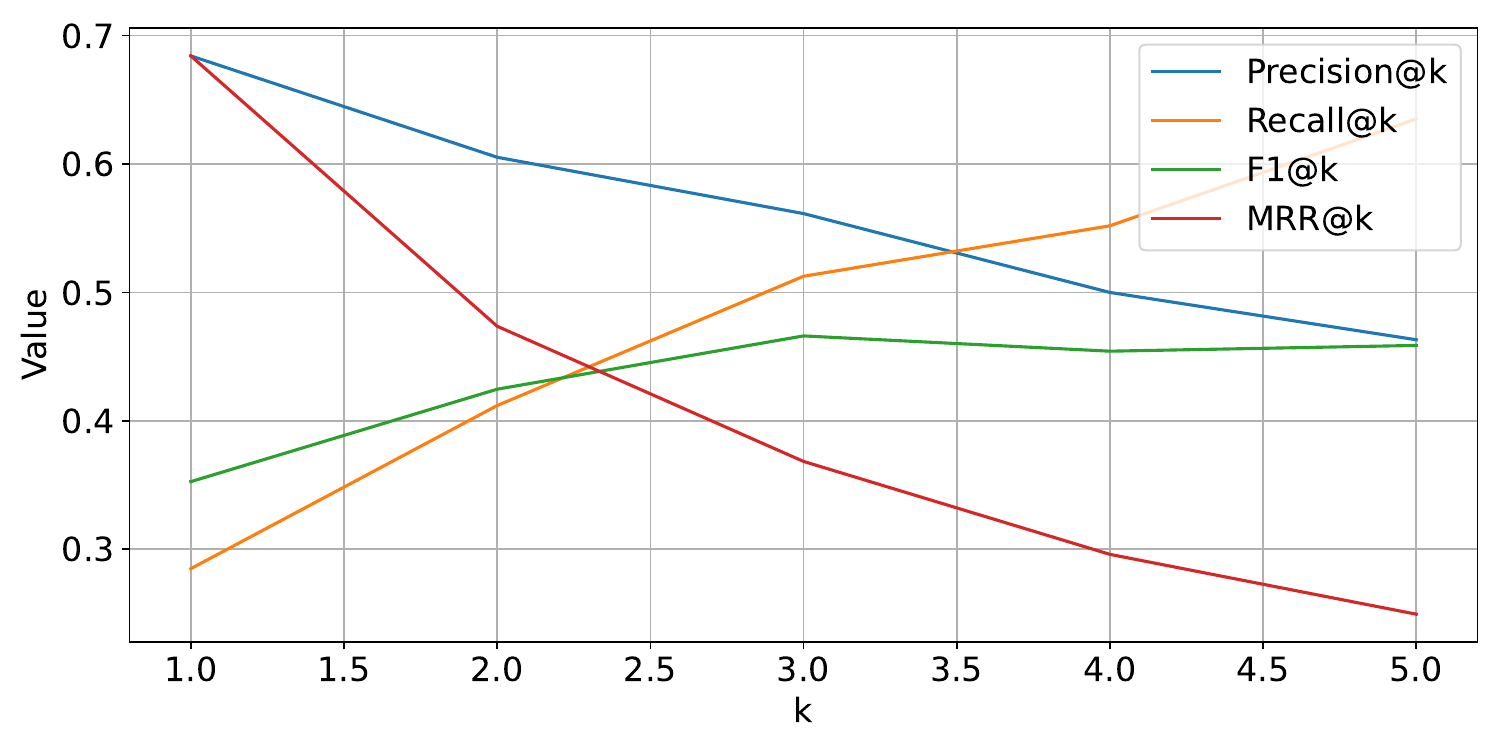}
    \caption{Results of Top-K \fixpatch selection via \enpassrate.}
    \label{fig:rq3_top_k}
\end{figure}

Figure~\ref{fig:rq3_top_k} shows precision@$K$ and recall@$K$ achieved with the \enpassratefull (\enpassrate) computed on the generated \brt{}s, $K$ increments from 1 to 5. 
We can see from Figure~\ref{fig:rq3_top_k} that \enpassrate allows for different \fixpatch selection strategies. 

When prioritizing maximum precision, selecting the top-ranked \fixpatch ($K=1$) correctly identifies a plausible \fixpatch from a pool of 20 candidates in 70\% of cases, achieving a precision of 0.7. 
This means that, on average, the correct \fixpatch is ranked first in 70\% of the cases (an MRR of 0.7). However, this approach yields a lower recall of 0.3.  

Alternatively, a more balanced approach, considering the top-3 ranked \fixpatch{}es ($K=3$), achieves a precision of 0.6 and a recall of 0.5 obtaining the highest F1-score among the different K values. 
Varying $K$ between 1 and 5 reveals a trade-off: precision decreases from 0.7 to 0.5, while recall increases from 0.3 to 0.6. 

Overall, Top-K selection via \enpassrate allows for only the most likely plausible \fixpatch{}es to be presented to the \autopr user\Comment{/developer\samcheng{user in this paper is developer}}, minimizing the need to review incorrect \fixpatch{}es.

\subsubsection{Threshold-Based Selection}
\begin{figure}[t!]
    \centering
    \includegraphics[width=\columnwidth]{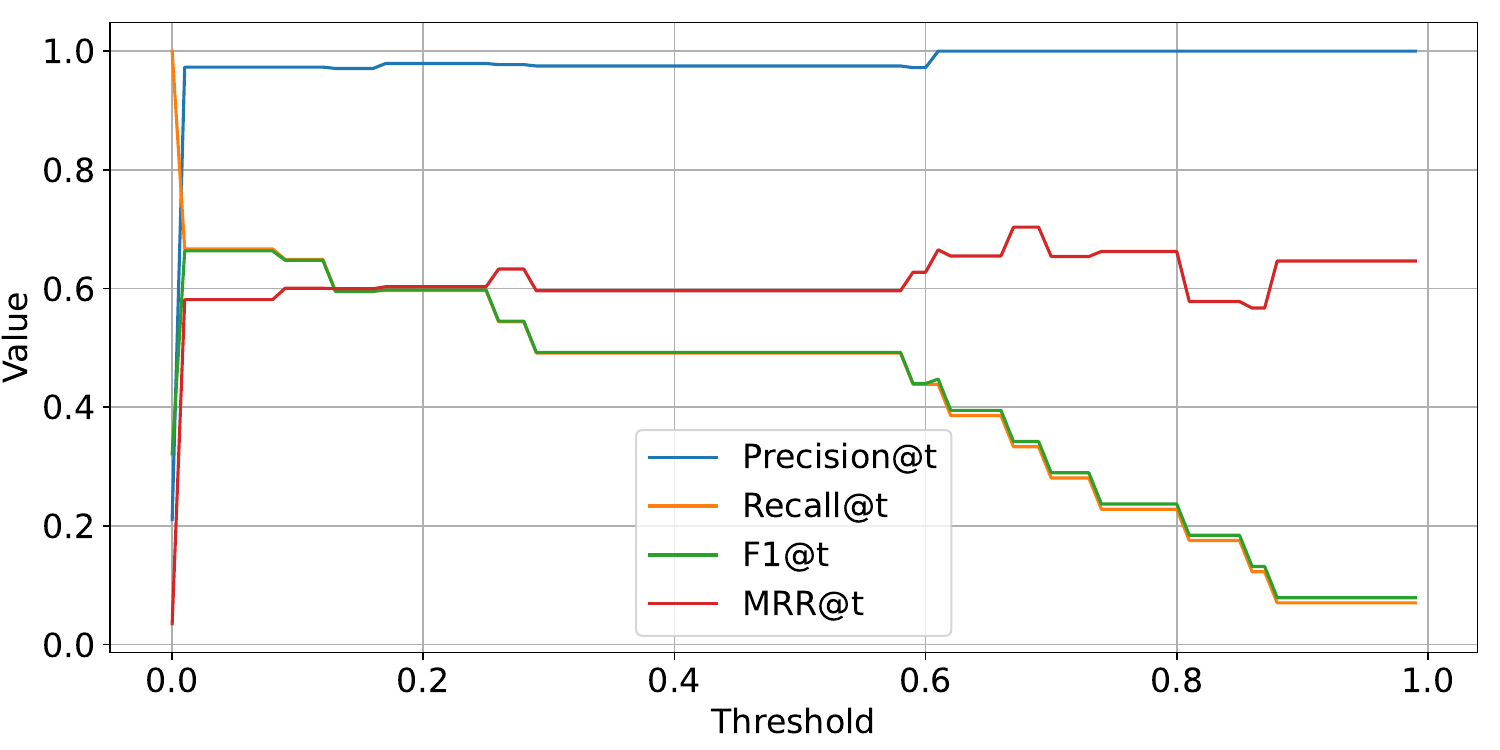}
    \caption{Results of threshold-based \fixpatch selection via \enpassrate.}
    \label{fig:rq3_threshold}
\end{figure}

Figure \ref{fig:rq3_threshold} illustrates the trade-off between precision and recall when varying the \enpassrate threshold for \fixpatch selection. 
As the \enpassrate threshold increases from 0 to 1 in increments of 0.01, precision increases from 0.3 to 1, while recall decreases from 1 to 0.1. 
A balance point is when the threshold is set to 0.1 (e.g., \fixpatch{}es passed at least 1-2 generated \brt{}s), where we achieve a precision of 0.9, and a recall of 0.6. This threshold also yields an MRR of 0.6, indicating that a correct \fixpatch is frequently ranked highly. 
Additionally, with an F1-score of 0.6, this threshold offers a good balance between precision and recall for \fixpatch selection.

These results demonstrate that the \enpassratefull (\enpassrate), derived from generated candidate \brt{}s, is an effective metric for selecting plausible fixes. 
While it exhibits good precision, the recall can be improved by generating more candidate fixes per bug.

\section{Threats to Validity}
\label{sec:threats}

While our study provides promising results for automated \brt generation in an industrial setting, several factors could potentially influence the validity of our findings.\Space{We categorize these threats into internal, external, and construct validity.}

\myparagraph{Internal Validity}
One potential threat\Space{ to internal validity} is the evaluation dataset size.
We study \brt generation on a relatively small dataset of 80 bugs from \google's internal issue tracking system (\S\ref{sec:empiricalstudy:datasetandconfig}). 
Although this dataset has been carefully curated and studied for prior work on \autopr~\cite{rondon2025passerine}, its size could limit the generalizability of our results to the broader spectrum of bugs encountered within \google.\Space{The specific distribution of bug types and affected projects within the dataset might not perfectly reflect the overall bug landscape.}
Another potential threat is implementation bias. 
Our adaptation of \libro may have introduced subtle differences that could affect its performance.
To mitigate such threat, we ensure each change we made to \libro improves its experiment performance in \google's environment. 
The difference in \llm{}s can also pose a threat.
To mitigate such threat, we consistently use the same \gemini models(s) for code generation for all evaluated techniques.
Finally, the inherent randomness in \llm{}s could influence performance of \llm-based techniques. 
To mitigate the threat from such randomness, we run each evaluated technique on the same input multiple times (\S\ref{sec:empiricalstudy:datasetandconfig}).

\myparagraph{External Validity}
A key threat to external validity is the generalizability of our findings to other industrial settings. 
Our study focuses exclusively on \google's internal development environment. 
While this provides valuable insights into a large-scale industrial setting, the specific tools, processes, and codebase characteristics may differ significantly from those in other companies. 
Therefore, the generalizability of our findings to other industrial settings requires further investigation.

\myparagraph{Construct Validity}
One potential threat\Space{ to construct validity} lies in our evaluation.
We use metrics like Candidate \brt{}s and Plausible \brt{}s from prior work~\cite{mundler2024swt} to measure \brt generation success. 
However, these metrics may not fully capture all aspects of a\Space{ ``good''} \brt, such as its readability, maintainability, or ability to trigger subtle, hard-to-reproduce bugs. 
To mitigate this threat, we perform manual inspection on the plausible generated \brt{}s  (\S\ref{sec:results:brt}). 
Another threat to construct validity lies in our proposed metric \enpassrate (\S\ref{sec:empiricalstudy:rq3}) for assessing fix correctness in \S\ref{sec:empiricalstudy:rq3}. 
While \enpassrate shows promising results, it is inherently an indirect measure, and may not always perfectly correlate with the actual correctness of a fix as determined by human developers.

\section{Conclusion}
\label{sec:conclusion}

This paper investigates the potential for automated \brtfull (\brt) generation within a large-scale industrial setting, specifically at \google. 
We adapt and evaluate an existing \llm-based approach for \brt generation, specifically \libro \cite{kang2023large}, to function within \google's complex internal development environment.
We also introduce our \llm-agent-based \brt generation approach, designed to work with \google's development environment. 
\tool, our \brt generation technique that is built on top of a \gemini model fine-tuned on \google's codebase, significantly outperforms the adapted \libro built on top of the same \gemini model, demonstrating the efficacy of agent-based system for effective \brt generation.

Our empirical results show that \tool can generate plausible \brt{}s for more bugs (28\% compared to 10\% by \libro). 
These generated plausible \brt{}s can improve the performance of \passerine, \google's industrial-scale \autopr system, in generating plausible fixes for more bugs more efficiently. 
Finally, we show that candidate \brt{}s with \failtoany behavior can also be used to effectively rank and select plausible fixes generated by \autopr leveraging our proposed \enpassratefull metric.

These findings underscore the practical value of \brt{}s and highlight the importance of developing robust, industry-aware \brt generation techniques. 
We hope this research contributes to bridging the gap between academic research and industrial needs, offering a practical solution for improving software quality and accelerating bug resolution in large-scale, complex software ecosystems.

\balance
\bibliographystyle{ACM-Reference-Format}
\bibliography{references}

\end{document}